\documentclass[12pt]{article}
\usepackage{epsfig}
\begin{document}

\begin{center}

{\large \bf PHOTOPRODUCTION OF $J/\psi$ MESONS AT HIGH ENERGIES \\
 IN PARTON MODEL AND  $k_\top$-FACTORIZATION APPROACH}

\vspace{4mm}

{\sl Saleev V.A.}\footnote
{Email: saleev@ssu.samara.ru}\\[2mm]
Samara State University, 443011, Samara\\
and\\
Samara Municipal Nayanova
University, 443001, Samara\\
Russia
\end{center}



\begin{abstract}
We consider  $J/\psi$ meson photoproduction on protons at high
energies at the leading order in $\alpha_s$ using conventional
parton model and $k_\top$-factorization approach of QCD. It is
shown that in the both cases the colour singlet mechanism gives
correct description for experimental data from HERA for the total
cross section and  for the $J/\psi$ meson z-spectrum at realistic
values of a c-quark mass and meson wave function at the origin
$\Psi (0)$. At the same time our predictions for $p_\top$-spectrum
of $J/\psi$ meson and for $p_\top$ dependence of the spin
parameter $\alpha$ obtained in $k_\top$-factorization approach are
very different from the results obtained in conventional parton
model. Such a way the experimental study of a polarized $J/\psi$
meson production at the large $p_\top$ should be a direct test of
BFKL gluons.
\end{abstract}


\section{Introduction}
It is well known that in the processes of $J/\psi$ meson
photoproduction on protons at high energies the photon-gluon
fusion partonic subprocess dominates \cite{1}. In the framework of
general factorization approach of QCD the $J/\psi$ photoproduction
cross section depends on the gluon distribution function in a
proton, the hard amplitude of $c\bar c$-pair production as well as
the mechanism of a creation colorless final state with quantum
numbers of the $J/\psi$ meson. Such a way, we suppose that the
soft interactions in the initial state are described by
introducing a gluon distribution function, the hard partonic
amplitude is calculated using perturbative theory of QCD at order
in $\alpha_s(\mu^2)$, where $\mu\sim m_c$, and the soft process of
the $c\bar c$-pair transition into $J/\psi$ meson is described in
nonrelativistic approximation using series in the small parameters
$\alpha_s$ and $v$ (relative velocity of the quarks in $J/\psi$
meson). As it is statement in nonrelativistic QCD (NRQCD)
\cite{2}, there are colour singlet mechanism, where $c\bar c$-pair
is hardly produced in the colour singlet state, and colour octet
mechanism, where $c\bar c$-pair is produced in the color octet
state and at long distance it transforms into final colour singlet
state in soft process. However, as it was shown in paper \cite{3},
the data from HERA collider \cite{4} in the wide region of
$p_\top$ and $z$ may be described well in the framework of the
colour singlet model and not colour octet contribution is needed.
Based on the above mentioned result we will take into account in
our analysis only the colour singlet model contribution in the
$J/\psi$ meson photoproduction \cite{1}. We consider a role of a
proton gluon distribution function in the $J/\psi$ photoproduction
in the framework of the conventional parton model as well as in
the framework of the $k_\top$-factorization approach \cite{5}.
Last one is based on BFKL evolution equation \cite{6}, which takes
into account large terms proportional to $\log 1/x$ and $\log
\mu^2/\Lambda_{QCD}^2$ opposite to DGLAP evolution equation
\cite{7}, where only large logarithmic  terms $\sim \log
\mu^2/\Lambda_{QCD}^2$ are taken into account. In the process of
the $J/\psi$ meson photoproduction one has $x\sim \mu^2/s$ and
$\mu^2\sim M^2$, where $M$ is the $J/\psi$ meson mass, $s$ is the
square of a total energy of colliding particles in center of mass
frame.

\section{The cross section for $\gamma p\to J/\psi X$
in $k_\top$-factorization approach}
Nowadays, there are two approaches in calculating of $J/\psi$
meson or heavy quark production cross sections at high energies.
In the conventional parton model \cite{8} it is suggested that
hadronic cross section
 $\sigma (\gamma p\to J/\psi X,s)$ and relevant partonic cross
section $\hat \sigma (\gamma g\to J/\psi
g,\hat s)$ are connected as follows:
\begin{equation}
\sigma^{PM}(\gamma p\to J/\psi X,s)=
\int dx\hat\sigma (\gamma g\to J/\psi g, \hat s) G(x,\mu^2),
\end{equation}
where $\hat s=x s$,
 $G(x,\mu^2)$ is the collinear gluon distribution function
in a proton,  $x$ is the fraction of a proton momentum, $\mu^2$ is
the typical scale of a hard process. The $\mu^2$ dependence of the
gluon distribution $G(x,\mu^2)$ is described by DGLAP evolution
equation \cite{7}. In the region of very high $s$ one has $x<<1$.
This fact  leads to BFKL evolution equation \cite{6} for the
unintegrated gluon distribution function $\Phi (x,\bf
q_\top^2,\mu^2)$, where $\bf q_\top^2$ is the gluon virtuality.
The unintegrated gluon distribution function can be related to the
conventional gluon distribution by
\begin{equation}
xG(x,\mu^2)=\int_0^{\mu^2}\Phi (x,{\bf q}_\top^2,\mu^2)d{\bf q}_\top^2.
\end{equation}
The gluon 4-momentum is presented as follows
$$ q=x p_N+q_{\top},$$
where $q_{\top}=(0,{\bf q}_{\top},0)$, $p_N=(E_N,0,0,|{\bf p}_N|)$
and $q^2= q^2_{\top}=-{\bf q_{\top}}^2$. So called BFKL gluon is
off mass-shell and it has polarization vector along its transverse
momentum such as $\varepsilon^{\mu}=q_{\top}^{\mu}/|\bf
q_{\top}|$. In $k_\top$-factorization approach hadronic and
partonic cross sections are related by the following condition:
\begin{equation}
\sigma (\gamma p\to J/\psi X)=
\int \frac{dx}{x}\int_0^{\mu^2}d {\bf q}^2_{\top}\int \frac{d\phi}{2 \pi}
\Phi (x,{\bf q}^2_{\top},\mu^2)
\hat \sigma (\gamma g^*\to J/\psi g,\hat s, {\bf q}^2_{\top}),
\end{equation}
where $\hat \sigma (\gamma g^*\to J/\psi g)$ is
$J/\psi$ meson photoproduction on BFKL gluon, $\phi$
is azimuthal angle in the transverse $XY$ plane between
vector ${\bf q}_{\top}$ and fixed  $OX$ axis.


\section{Unintegrated gluon distribution function}
At the present time an exact form of the unintegrated gluon
distribution  $\Phi (x, {\bf q}^2_{\top}, \mu^2)$ is unknown yet
because the relevant experimental analysis of the experimental
data has never been carried out. There are several theoretical
approximations for $\Phi (x, {\bf q}^2_{\top}, \mu^2)$, which are
based on solving BFKL evolution equation \cite{9a,9b,9c,9}. In the
region of very small $x\le 0.01$ and moderate ${\bf q}^2_{\top}$
($\sim 10$ GeV$^2$), which is relevant for $J/\Psi$
photoproduction at HERA, all parameterizations \cite{9a,9b,9} are
approximately coincide (see, for example, discussions in \cite{9d}
and \cite{9e}). For our purposes, we will use the unintegrated
gluon distribution, which was obtained in the paper \cite{9}. The
proposed method lies upon a straightforward perturbative solution
of the BFKL equation where the collinear gluon density
 $G(x,\mu^2)$ is used as the boundary condition in the integral form
(2). Technically, the unintegrated gluon distribution is
calculated as a convolution of collinear gluon distribution
$G(x,\mu^2)$ with universal weight factors:
\begin{equation}
\Phi (x,{\bf q}^2_{\top},\mu^2)=\int_x^1 {\cal G}(\eta, {\bf q}^2_{\top},
\mu^2)\frac{x}{\eta} G(\frac{x}{\eta},\mu^2)d \eta,
\end{equation}
where
\begin{eqnarray}
&&{\cal G}(\eta, {\bf q}^2_{\top},\mu^2)=
\frac{\bar\alpha_s}{\eta {\bf q}^2_{\top}}
J_0 \left[2\sqrt{\bar\alpha_s\ln (\frac{1}{\eta})\ln
(\frac{\mu^2}{{\bf q}^2_{\top}})}\right ],
\mbox{ at }{\bf q}^2_{\top}\le  \mu^2 \\
&&{\cal G}(\eta, {\bf q}^2_{\top}, \mu^2)=
\frac{\bar\alpha_s}{\eta {\bf q}^2_{\top}} I_0
\left[2\sqrt{\bar\alpha_s\ln (\frac{1}{\eta})\ln (\frac{{\bf
q}^2_{\top}}{\mu^2})}\right ], \mbox{ at }{\bf q}^2_{\top}> \mu^2,
\end{eqnarray}
$J_0$ and $I_0$ stand for Bessel functions (of real and imaginary
arguments, respectively), and $\bar
\alpha_s=\displaystyle{\frac{\alpha_s}{3\pi}}$. As a input
function $G(x,\mu^2)$ in (5,6) we use the standard GRV
parameterization  \cite{10}. To test the method of calculation for
$\Phi (x, {\bf q}^2_{\top}, \mu^2)$ we compare input collinear
gluon distribution $G(x,\mu^2)$ and collinear gluon distribution
$\tilde G(x,\mu^2)$, which is obtained using formula (2) from the
unintegrated distribution $\Phi (x, {\bf q}^2_{\top}, \mu^2)$
after integration over ${\bf q}^2_{\top}$. In the Figure 1 the
result of our calculation for the ratio $R_1(x, \mu^2)=\tilde
G(x,\mu^2)/G(x,\mu^2)$ is shown. It is visible, that at  $x<0.1$
the ratio  $R_1$ does not differ from 1 more than 1-2 \%. Note,
that in a process of the $J/\psi$  photoproduction on protons
$x\sim \displaystyle{\frac{M}{\sqrt{s}}}$ ($M$ is the mass of
$J/\psi$ meson), and one has $x\sim 0.03$ at $\sqrt {s}=100$ GeV.

\section{Amplitude for $\gamma g\to J/\psi g$ process}
There are six Feynman diagrams (Figure 2) which describere
partonic process $\gamma g \to J/\psi g$ at the leading order in
$\alpha_s$ and $\alpha$. In the framework of the colour singlet
model and nonrelativistic approximation the production of $J/\psi$
meson is considered as the production of a quark-antiquark system
in the colour singlet state with orbital momentum $L=0$ and spin
momentum $S=1$. The binding energy and relative momentum of quarks
in the $J/\psi$ are neglected. Such a way $M=2 m_c$ and
$p_c=p_{\bar c}=\displaystyle{\frac{p}{2}}$, where $p$ is
4-momentum of the $J/\psi$,  $p_c$ and $p_{\bar c}$ are 4-momenta
of quark and antiquark. Taking into account the formalism of the
projection operator \cite{11} the amplitude of the process $\gamma
g\to J/\psi g$ may be obtained from the amplitude of the process
$\gamma g\to \bar c c g$ after replacement:
\begin{equation}
V^i(p_{\bar c})\bar U^j(p_c) \rightarrow \frac{\Psi(0)}{2\sqrt{M}}\hat
\varepsilon (p)(\hat p+M)\frac{\delta^{ij}}{\sqrt{3}},
\end{equation}
where $\hat \varepsilon (p)=\varepsilon_{\mu}(p)\gamma^{\mu}$,
$\varepsilon_{\mu}(p)$ is 4-vector of the $J/\psi$ polarization,
$\displaystyle{\frac{\delta^{ij}}{\sqrt{3}}}$ is the color factor,
$\Psi(0)$ is the nonrelativistic meson wave function at the
origin. The matrix elements of the process $\gamma g^*\to J/\psi
g$ may be presented as follows:
\begin{equation}
M_i=KC^{ab}\varepsilon_{\alpha}(k_1)\varepsilon^a_{\mu}(q)
\varepsilon^b_{\beta}(k_2)\varepsilon_{\nu}(p) M_i^{\alpha\beta\mu\nu},
\end{equation}
\begin{eqnarray}
&&M_1^{\alpha\beta\mu\nu}=
\mbox{Tr}\left [\gamma^{\nu}(\hat p+M)\gamma^{\alpha}\frac{\hat
p_c-\hat k_1+m_c}{(p_c-k_1)^2-m_c^2}\gamma^{\mu}\frac{-\hat p_{\bar c}-\hat
k_2+m_c}{(p_{\bar c}+k_2)^2-m_c^2}\gamma^{\beta}\right ]
\\
&&M_2^{\alpha\beta\mu\nu}=
\mbox{Tr}\left [\gamma^{\nu}(\hat p+M)\gamma^{\beta}\frac{\hat
p_c+\hat k_2+m_c}{(p_c+k_2)^2-m_c^2}\gamma^{\alpha}\frac{\hat q -\hat
p_{\bar c}+m_c}{(q-p_{\bar c})^2-m_c^2}\gamma^{\mu}\right ]
\\
&&M_3^{\alpha\beta\mu\nu}=
\mbox{Tr}\left [\gamma^{\nu}(\hat p+M)\gamma^{\alpha}\frac{\hat
p_c-\hat k_1+m_c}{(p_c-k_1)^2-m_c^2}\gamma^{\beta}\frac{\hat q-\hat
p_{\bar c}+m_c}{(q-p_{\bar c})^2-m_c^2}\gamma^{\mu}\right ]
\\
&&M_4^{\alpha\beta\mu\nu}=
\mbox{Tr}\left [\gamma^{\nu}(\hat p+M)\gamma^{\mu}\frac{\hat
p_c-\hat q+m_c}{(p_c-q)^2-m_c^2}\gamma^{\alpha}\frac{-\hat p_{\bar c}-\hat
k_2+m_c}{(k_2+p_{\bar c})^2-m_c^2}\gamma^{\beta}\right ]
\\
&&M_5^{\alpha\beta\mu\nu}=
\mbox{Tr}\left [\gamma^{\nu}(\hat p+M)\gamma^{\beta}\frac{\hat
p_c+\hat k_2+m_c}{(p_c+k_2)^2-m_c^2}\gamma^{\mu}\frac{\hat k_1 -\hat
p_{\bar c}+m_c}{(k_1-p_{\bar c})^2-m_c^2}\gamma^{\alpha}\right ]
\\
&&M_6^{\alpha\beta\mu\nu}=
\mbox{Tr}\left [\gamma^{\nu}(\hat p+M)\gamma^{\mu}\frac{\hat
p_c-\hat q+m_c}{(p_c-q)^2-m_c^2}\gamma^{\beta}\frac{\hat k_1-\hat
p_{\bar c}+m_c}{(k_1-p_{\bar c})^2-m_c^2}\gamma^{\alpha}\right ]
\end{eqnarray}
where  $k_1$ is the 4-momentum of the photon,  $q$ is 4-momentum
of the initial gluon,
$k_2$ is the  4-momentum of the final gluon,
$$ K=e_ceg_s^2\frac{\Psi(0)}{2\sqrt{M}},
\qquad C^{ab}=\frac{1}{\sqrt{3}}\mbox{Tr}[T^aT^b],\quad
e_c=\frac{2}{3},\quad e=\sqrt{4\pi\alpha},\quad
g_s=\sqrt{4\pi\alpha_s}.$$ The summation on photon, $J/\psi$ meson
and final gluon polarizations is carried out by covariant
formulae:
\begin{eqnarray}
&&\sum_{spin}\varepsilon_{\alpha}(k_1)\varepsilon_{\beta}(k_1)
=-g_{\alpha\beta},\\
&&\sum_{spin}\varepsilon_{\alpha}(k_2)\varepsilon_{\beta}(k_2)
=-g_{\alpha\beta},\\
&&\sum_{spin}\varepsilon_{\mu}(p)\varepsilon_{\nu}(p)
=-g_{\mu\nu}+\frac{p_{\mu}p_{\nu}}{M^2}.
\end{eqnarray}
In case of the initial BFKL gluon we use the following
prescription
\begin{equation}
\sum_{spin}\varepsilon_{\mu}(q)\varepsilon_{\nu}(q)=
\frac{q_{\top\mu}q_{\top\nu}}{{\bf q}^2_{\top}}.
\end{equation}
For studing  $J/\psi$ polarized photoproduction
we introduce the 4-vector of the longitudinal polarization
as follows
\begin{equation}
\varepsilon_L^{\mu}(p)=\frac{p^{\mu}}{M}-\frac{Mp_N^{\mu}}{(pp_N)}.
\end{equation}
In the limit of $s>>M^2$ the polarization 4-vector satisfies usual
conditions $(\varepsilon_L\varepsilon_L)=-1$, $(\varepsilon_L
p)=0$.

Traditionally for a description of quarkonium photoproduction
processes the invariant variable $z=(pp_N)/(k_1p_N)$ is used.
In the rest frame of the proton one has
$z=E_{\psi}/E_{\gamma}$.
In the $k_\top$-factorization approach the differential on
$p_{\top}$ and $z$ cross section of the $J/\psi$ photoproduction
may be written as follows
\begin{equation}
\frac{d\sigma}{dp^2_{\top}dz}=\frac{1}{z(1-z)}\int_0^{2\pi}
\frac{d\phi}{2\pi}\int_0^{\mu^2} d{\bf q}^2_{\top}\Phi (x,{\bf
q}^2_{\top},{\mu}^2) \frac{\overline{|M|^2}}{16\pi(xs)^2}.
\end{equation}
The numerical calculation is performed in the photon and proton
center of mass frame where
$$p_N=\frac{\sqrt{s}}{2}(1,0,0,1),\quad
k_1=\frac{\sqrt{s}}{2}(1,0,0,-1),$$
$$p=(E,{\bf p}_{\top},p_{\parallel}),\quad q=(\frac{\sqrt{2}}{2}x,
{\bf q}_{\top},\frac{\sqrt{s}}{2}x).$$
Here we take into account that the $J/\psi$ momentum
$\bf p$ lies in  $XZ$ plane and $({\bf
q}_{\top} {\bf p}_{\top})=|{\bf p}_{\top}| |{\bf q}_{\top}|\cos
(\phi)$.
The analytical calculation of the
$\overline{|M|^2}$ is performed with help of REDUCE package
and results are saved in the FORTRAN codes as a function of
$\hat s=(k_1+q)^2$,
$\hat t=(p-k_1)^2$, $\hat u=(p-q)^2$, ${\bf p}_\top^2$, ${\bf
q}^2_{\top}$ and  $\cos (\phi)$.
We directly have tested that
\begin{equation}
\lim_{{{\bf q}^2_{\top}\to 0}}\int_0^{2\pi}\frac{d\phi}{2\pi}
\overline{|M|^2}= \overline{|M_{PM}|^2},
\end{equation}
where ${\bf p}_\top^2= \displaystyle{\frac{\hat t \hat u}{\hat
s}}$ in the $|M|^2$ and $\overline{|M_{PM}|^2}$ is the square of
the amplitude in the conventional parton model \cite{3}. In the
limit of ${\bf q}^2_{\top}=0$ from formula (20) it is easy to find
the differential cross section in the parton model too:
\begin{equation}
\frac{d\sigma^{PM}}{dp^2_{\top}dz}=
\frac{\overline{|M_{PM}|^2}xG(x,\mu^2)}{16\pi(xs)^2z(1-z)}.
\end{equation}
However, making calculations in the parton model we use formula
(20), where integration over ${\bf q}^2_{\top}$ and $\phi$ is
performed numerically, instead of (22). This method fixes  common
normalization factor for both approaches and gives direct
opportunity to study effects connected with virtuality of the
initial BFKL gluon in the partonic amplitude.

\section{Results and discussion}
After we fixed selection of the gluon distribution function
$G(x,\mu^2)$ there are two parameters only, which values determine
the common normalization factor of the cross section under
consideration: $\Psi(0)$ and $m_c$. The value of the  $J/\psi$
meson wave function at the origin may be calculated in a potential
model or obtained from experimental well known decay width
$\Gamma(J/\psi\to \mu^+\mu^-)$. In our calculation we used the
following choice $|\Psi(0)|^2=0.0876$ GeV$^3$ as the same as in
Ref. \cite{3}. Concerning a charmed quark mass, the situation is
not clear up to the end. From one hand, in the nonrelativistic
approximation one has $m_c=\displaystyle{\frac{M}{2}}$, but there
are many evidences to take smaller value of a c-quark mass in the
amplitude of a hard process, for example $m_c=1.4$ GeV. Taking
into consideration above mentioned we perform calculations as at
$m_c=1.4$ GeV as well as at $m_c=1.55$  GeV. The cinematic region
under consideration is determined by the following conditions:
$0.4<z<0.9$ and $p_\top^2>1$ GeV$^2$, which correspond H1 and ZEUS
Collaborations data \cite{4}. We assume that the contribution of
the colour octet mechanism is large at the $z>0.9$ only. In the
region of the small values of $z<0.2$  the contribution of the
resolved photon processes \cite{55} as well as the charm
excitation processes \cite{66} may be large too. All of these
contributions are not in our consideration.

Figure 3 shows the ratio
\begin{equation}
R_2(\hat s,{\bf q}_\top^2 )=\frac{\hat\sigma (\hat s, {\bf
q}_\top^2)} {\hat \sigma(\hat s,0)},
\end{equation}
as a function of $\bf q_\top^2$ at the different $\hat s$. As well
as  it was necessary to expect ratio (23) decrease with growth of
${\bf q}_\top^2$ and the faster than less the value of $\hat s$.

Figures 4--6 show our results which were obtained as in the
conventional parton model as well as in the $k_\top$-factorization
approach at two values of a charmed quark mass. The dependence of the
results on selection of a hard scale parameter $\mu$ is much less
than the dependence on selection of a c-quark mass. We put $\mu^2=M^2+
\bf p_\top^2$ in a gluon distribution function and in a running
constant $\alpha_s(\mu^2)$.

Figure 4 shows the dependence of the total $J/\psi$ photoproduction
cross section on $\sqrt s$. It is visible that the difference
between the parton model prediction and the result of
$k_\top$-factorization approach is much less than between the results
obtained at different values of c-quark mass in the both models.
At the $m_c=1.55$ GeV the obtained cross sections in 1.5 --2 times
are less than experimental data \cite{4}, but at the
$m_c=1.4$ GeV the theoretical curves lie even a shade higher of the
experimental points.

There are not contradictions between the theoretical predictions
and data for the $z$-spectrum of the $J/\psi$ mesons.
Figure 5 shows that the experimental points lie inside the
theoretical corridor as in the parton model as in the
$k_\top$-factorization approach.

 The count of a transverse momentum of the BFKL gluons in the
$k_\top$-factorization approach results in a flattening of the
$p_\top$-spectrum of the $J/\psi$ as contrasted to by predictions
of parton model. For the first time on this effect was indicated
in the article \cite{12}, and later in the articles \cite{9e,14}.
Figure 6 shows the result of our calculation for the
$p_\top$-spectrum of the $J/\psi$ mesons. Using the
$k_\top$-factorization approach we have obtained the more hard
$p_\top$-spectrum of the  $J/\psi$ than it have been predicted in
the LO parton model. It is visible that at large values of
$p_\top$ only the $k_\top$-factorization approach gives correct
description of the data \cite{4}. However, it is impossible to
consider this visible effect as the direct indication on a
nontrivial developments of the small-x physics. In the article
\cite{3} was shown that calculation in the NLO approximation gives
more hard $p_\top$-spectrum of the $J/\psi$ meson too, which one
will be agreed the data at the large $p_\top$.

As it was mentioned above, the main difference between
the $k_\top$-factorization approach and the conventional parton
model is nontrivial polarization of BFKL gluon. It is obviously,
that such spin condition of the initial gluon should
result in observed spin effects during birth
of the polarized $J/\psi$ meson. We have performed calculations
for the spin parameter $\alpha$, as a function $z$ or $p_\top$
in the conventional parton model and in the
$k_\top$-factorization approach :
\begin{equation}
\alpha (z)=\frac {\frac{d\sigma_{tot}}{dz} -3\frac{d\sigma_L}{dz}}
{\frac{d\sigma_{tot}}{dz}+\frac{d\sigma_L}{dz}},
\end{equation}
\begin{equation}
\alpha (p_\top)=\frac { \frac{d\sigma_{tot}}{dp_{\top}}
-3\frac{d\sigma_L}{dp_\top}}
{\frac{d\sigma_{tot}}{dp_\top}+\frac{d\sigma_L}{dp_\top}}
\end{equation}
Here  $\sigma_{tot}=\sigma_L+\sigma_T$ is the total $J/\psi$
production cross section, $\sigma_L$ is the production cross
section for the longitudinal polarized $J/\psi$ mesons, $\sigma_T$
is the production cross section for the transverse polarized
$J/\psi$ mesons. The parameter $\alpha$ controls the angle
distribution for leptons in the decay $J/\psi \to l^+l^-$ in the
$J/\psi$ meson rest frame:
\begin{equation}
\frac{d\Gamma}{d \cos(\theta)}\sim 1+\alpha\cos(\theta).
\end{equation}
Figure 7 shows the parameter $\alpha (z)$, which is calculated in
the parton model (curve 2) and in the $k_\top$-factorization
approach (curve 1). We see that both curves lie near zero at
$z<0.8$ and increase at $z>0.8$. The large difference between
predictions is visible only at $z>0.9$ where our consideration is
not adequate. Let's remark, that parameter is gentle depends on
mass of a charmed quark and we demonstrate here only outcomes
obtained at the $m_c=1.55$.

For the parameter $\alpha (p_\top)$ we have found strongly
opposite predictions in the parton model and in the
$k_\top$-factorization approach, as it is visible in Figure 8. The
parton model predicts that  $J/\psi$ mesons should have transverse
polarizations at the large $p_\top$ ($\alpha(p_\top)=0.6$ at
$p_\top^2=20$ GeV$^2$), but  $k_\top$-factorization approach
predicts that $J/\psi$ mesons should be longitudinally polarized (
$\alpha(p_\top)=-0.4$ at $p_\top^2=20$ GeV$^2$). Nowadays, a
result of NLO parton model calculation in the case of the
polarized $J/\psi$ meson photoproduction is unknown. It should be
an interesting subject of future investigations. If the count of
the NLO corrections will not change predictions of the LO parton
model for $\alpha (p_\top)$, the experimental measurement of this
spin effect will be a direct signal about BFKL gluon dynamics.

Nowadays, the experimental data on $J/\Psi$ polarization in
photoproduction at large $p_\top$ are absent. However there are
similar data from CDF Collaboration \cite{16}, where $J/\psi$ and
$\psi'$ $p_\top$-spectra and polarizations have been measured.
Opposite the case of $J/\psi$ photoproduction, the hadroproduction
dada needs to take into account the large colour-octet
contribution in order to explain $J/\psi$ and $\psi'$ production
at Tevatron in the conventional collinear parton model. The
relative weight of colour-octet contribution may be smaller if we
use $k_\top$-factorization approach, as it was shown recently in
\cite{17,18,19,20}. The predicted using collinear parton model
transverse polarization of $J/\psi$ at large $p_\top$ is not
supported by the CDF data, which can be roughly explain in
$k_\top$-factorization approach \cite{18}. In conclusion, the
number of theoretical uncertainties in the case of $J/\psi$ meson
hadroproduction is much more than in the case of photoproduction
and they need more complicated investigation, that is why the
future experimental analysis of $J/\psi$ photoproduction at THERA
will be clean check of the collinear parton model and
$k_\top$-factorization approach.

This work has been supported in part by the programme "Universities of
Russia -- Basic Researches" under Grant 02.01.003.



%

\newpage

\begin{figure}
\epsfig{figure=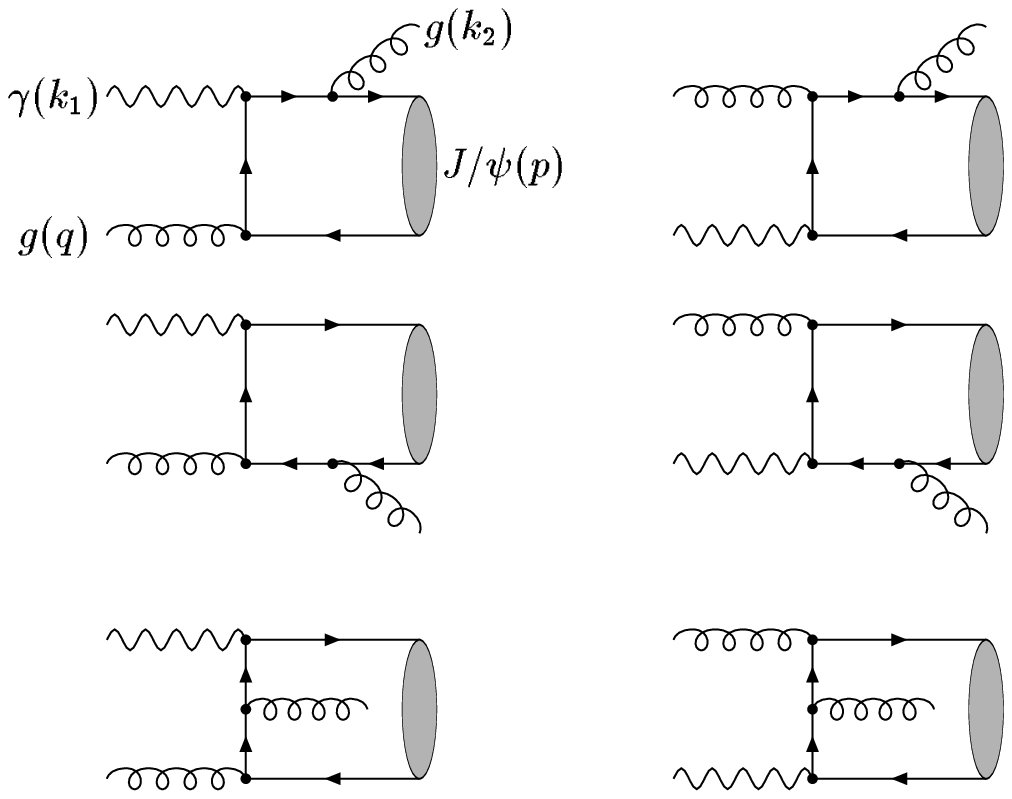,width=10cm} \caption{Diagrams used for
description partonic process $\gamma+g\to J/\psi+g$.}
\end{figure}

\vspace{15mm}

\begin{figure}[b]
\epsfig{figure=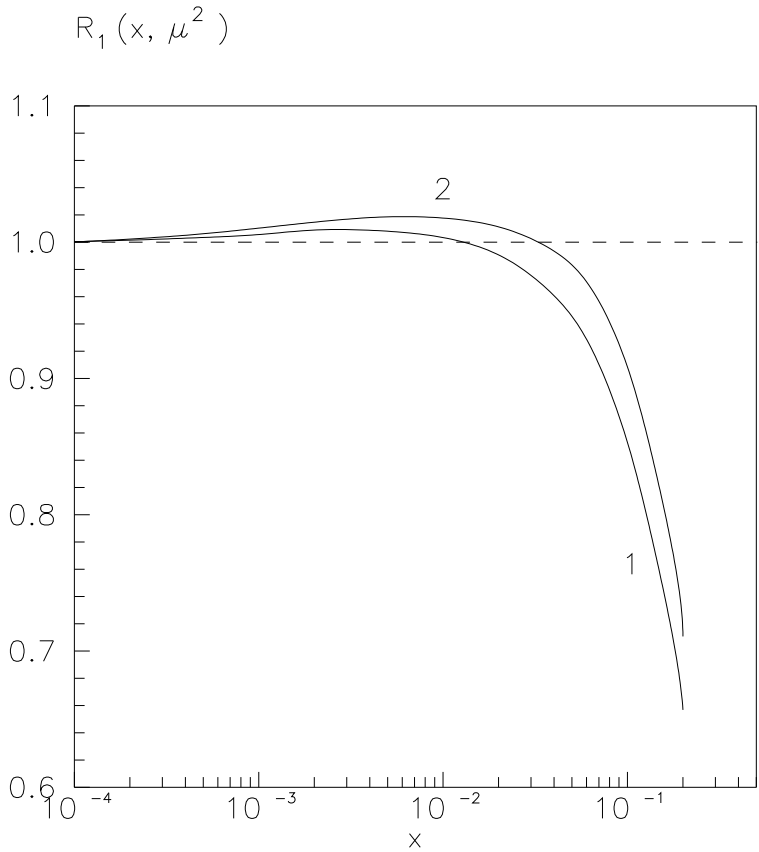,width=10cm} %
\caption{Ratio $R_1(x,\bf \mu^2)$ at $\bf \mu^2=10$ GeV$^2$ (curve
1) and $\bf \mu^2=100$ GeV$^2$ (curve 2).}
\end{figure}

\newpage

\begin{figure}
\epsfig{figure=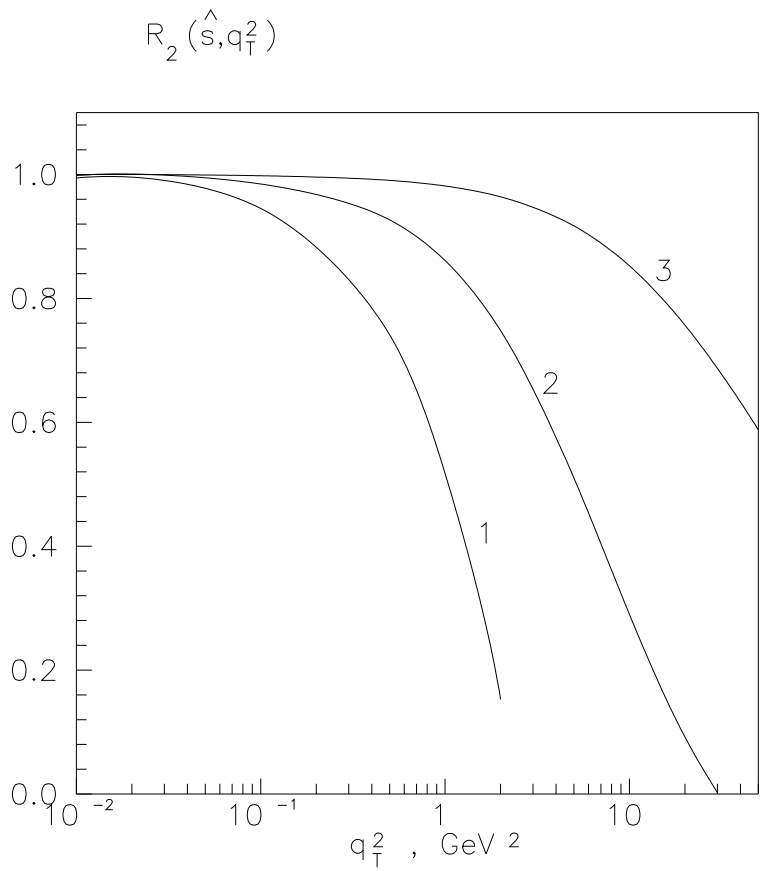,width=10cm} %
\caption{Ratio $R_2(\hat s,{\bf q}_{\top}^2)$ at $\hat s=15, 50,
100$ GeV$^2$ -- curves 1, 2, and 3, respectively.}
\end{figure}

\newpage

\begin{figure}
\epsfig{figure=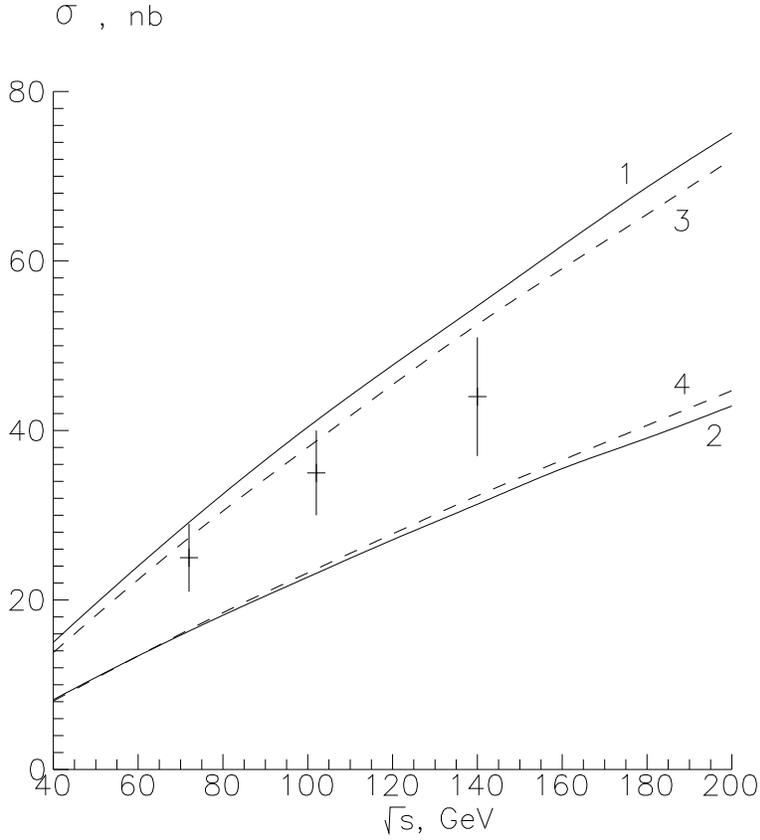,width=10cm} %
\caption{The
 $J/\psi$ photoproduction cross section as a function of $\sqrt s$
at $0.4<z<0.9$ and $p^2_{\top}>1$ GeV$^2$. The data from [4], the
curves 1 and 2 are obtained in the $k_\top$-factorization approach
at $m_c=1.4$ and $m_c=1.55$ GeV, the curves 3 and 4 are obtained
in the parton model at $m_c=1.4$ GeV and $m_c=1.55$ GeV.}
\end{figure}

\newpage
\begin{figure}
\epsfig{figure=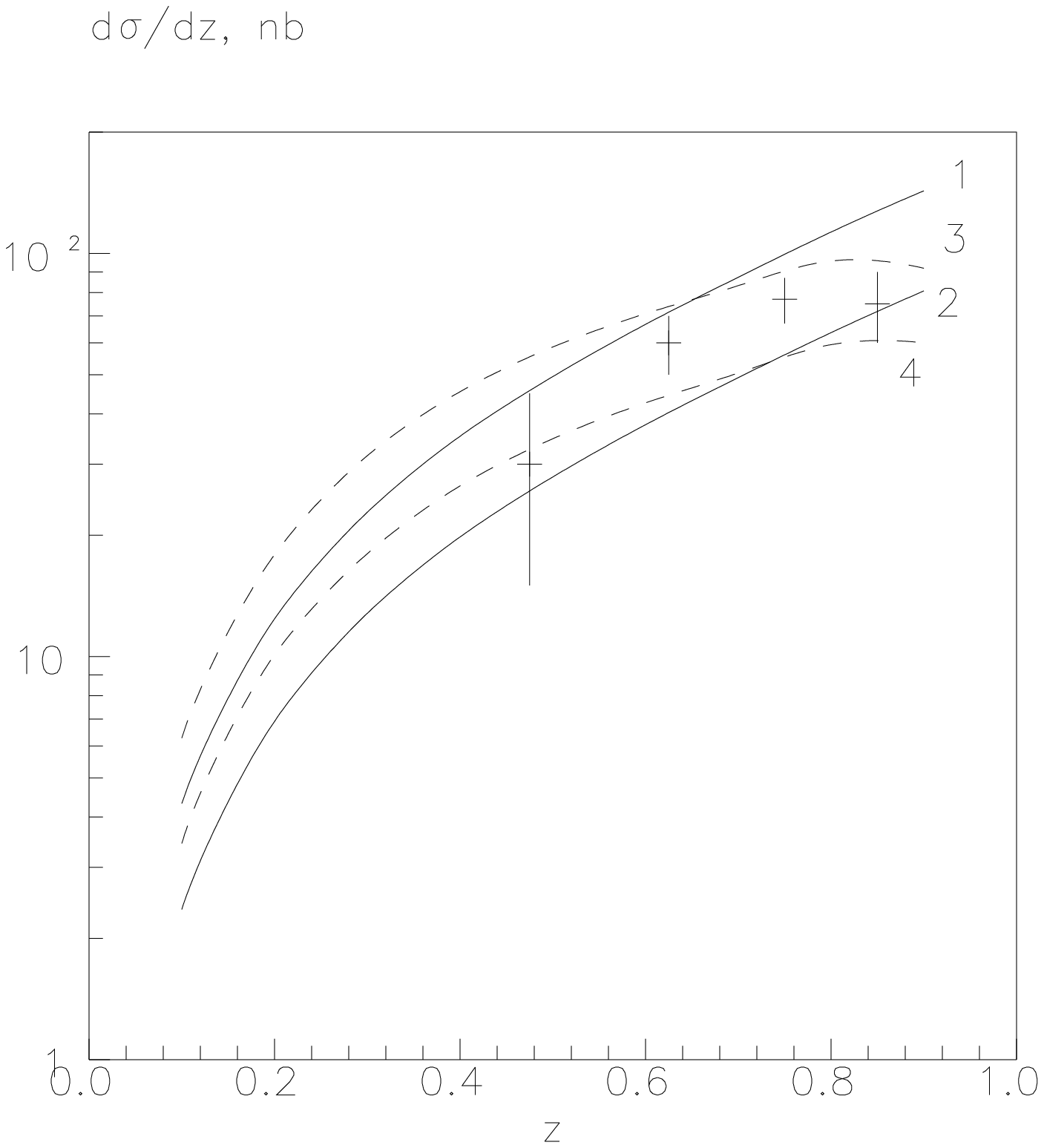,width=10cm} %
\caption{The
 $J/\psi$ spectrum on $z$ at $\sqrt s=100$ GeV
and $p^2_{\top}>1$ GeV$^2$.
The data from [4], the curves are the same as in Figure 4.}
\end{figure}

     \newpage
\begin{figure}
\epsfig{figure=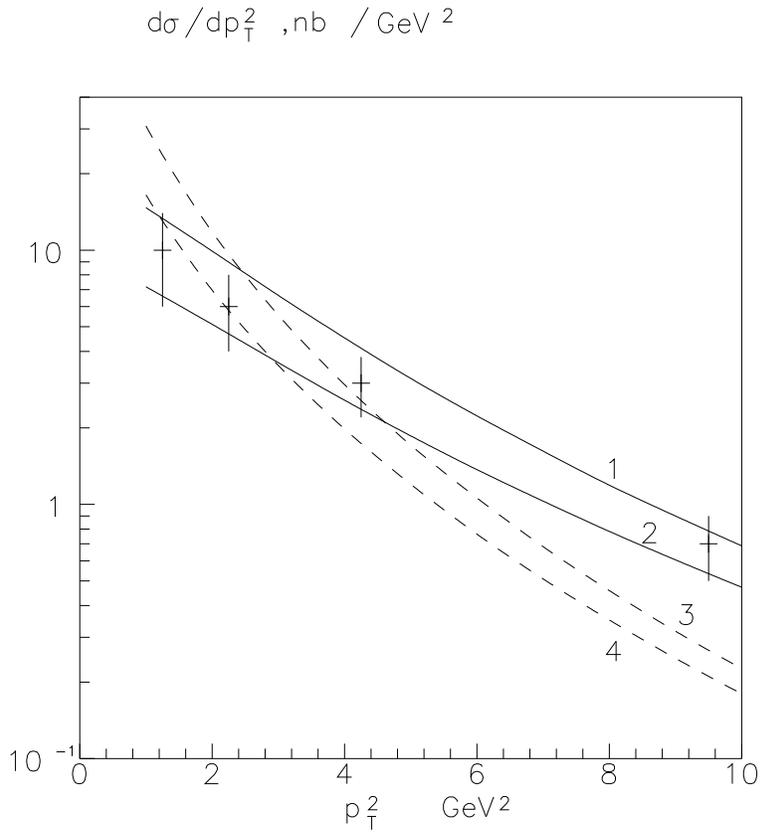,width=10cm} %
\caption{The  $J/\psi$ spectrum on $p^2_{\top}$ at $\sqrt s=100$
GeV and $0.4<z<0.9$. The data from [4], the curves are the same as
in Figure 4.}
\end{figure}

     \newpage
\begin{figure}
\epsfig{figure=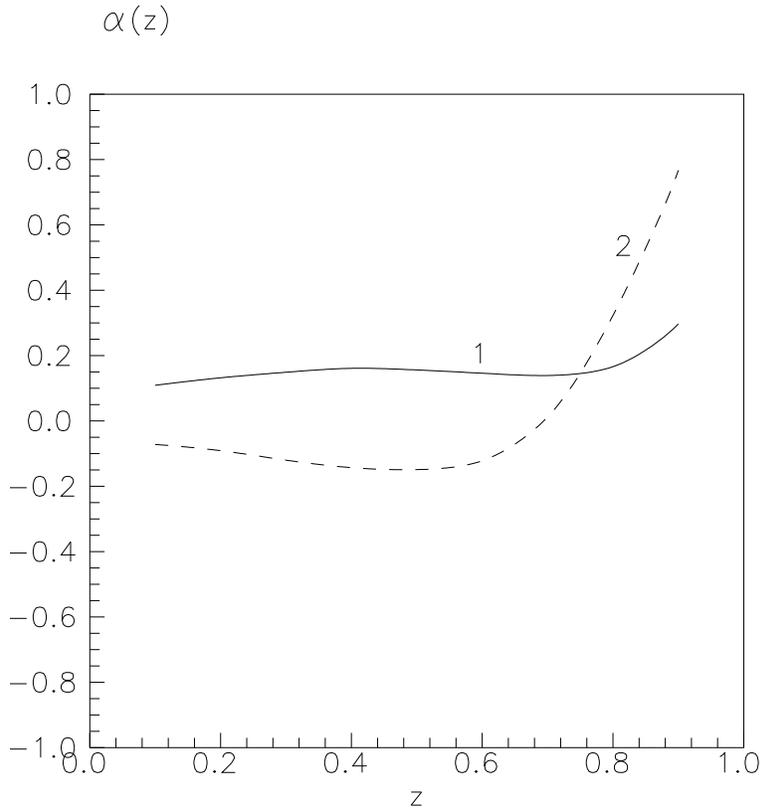,width=10cm} %
\caption{
The parameter $\alpha$ as a function of $z$ at $\sqrt s=100$ GeV,
 $p^2_{\top}>1$ GeV$^2$ and $m_c=1.55$ GeV.
The curve 1 is the result obtained in the $k_\top$-factorization
approach, the  curve 2 is the result obtained in the parton
model.}
\end{figure}

     \newpage
\begin{figure}
\epsfig{figure=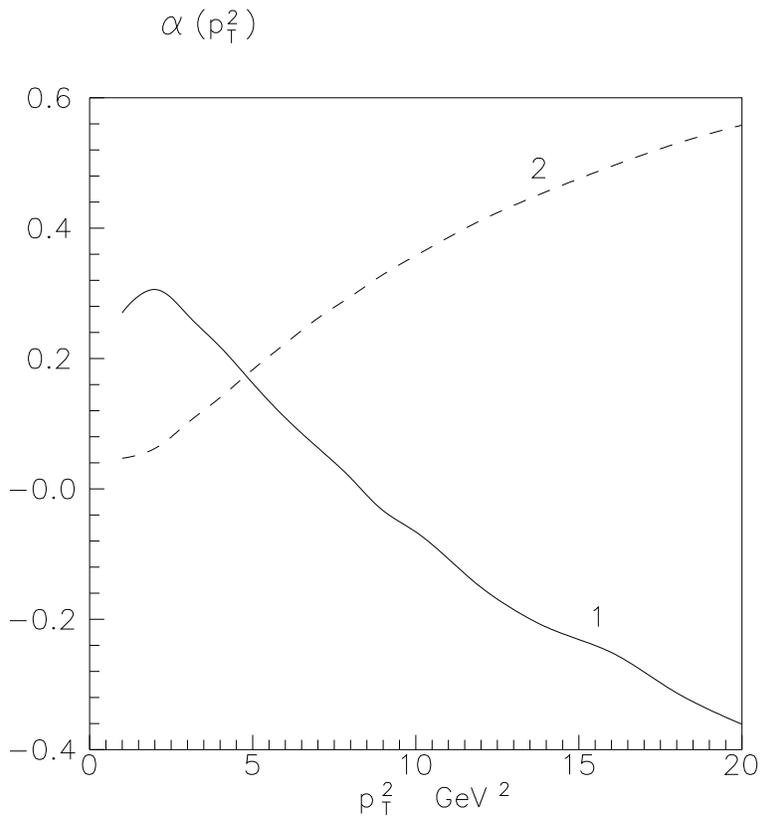,width=10cm} %
\caption{ The parameter $\alpha$ as a function of $p^2_{\top}$ at
$\sqrt s=100$ GeV, $0.4<z<0.9$ and $m_c=1.55$ GeV. The curves are
the same as in Figure 7.}

\end{figure}

\end{document}